\documentclass{article}

     \usepackage[final]{neurips_2020}

\usepackage[utf8]{inputenc} 
\usepackage[T1]{fontenc}    
\usepackage{hyperref}       
\usepackage{bibunits}
\usepackage{amsthm}
\usepackage{url}            
\usepackage{booktabs}       
\usepackage{amsfonts}       
\usepackage{nicefrac}       
\usepackage{microtype}      

\title{Biased Programmers? Or Biased Data? A Field Experiment in Operationalizing AI Ethics}

\author{%
Bo Cowgill and Fabrizio Dell\textquoteright Acqua  \\
  Graduate School of Business, Columbia University\\
  \texttt{\{bo.cowgill,fdellacqua21\}@gsb.columbia.edu,}
  \And
 Samuel Deng, Daniel Hsu, Nakul Verma, Augustin Chaintreau \\
  Department of Computer Science, Columbia University\\
  \texttt{samuel.d@columbia.edu,\{djhsu,verma,augustin\}@cs.columbia.edu}
  }

\begin{document}

\maketitle
\begin{abstract}
Why do biased predictions arise? What interventions can prevent them? We evaluate 8.2 million algorithmic predictions of math performance from $\approx400$ AI engineers, each of whom developed an algorithm under a randomly assigned experimental condition. Our treatment arms modified programmers' incentives, training data, awareness, and/or technical knowledge of AI ethics. We then assess out-of-sample predictions from their algorithms using randomized audit manipulations of algorithm inputs and ground-truth math performance for 20K subjects. We find that biased predictions are mostly caused by biased training data. However, one-third of the benefit of better training data comes through a novel economic mechanism: Engineers exert greater effort and are more responsive to incentives when given better training data. We also assess how performance varies with programmers' demographic characteristics, and their performance on a psychological test of implicit bias (IAT) concerning gender and careers. We find no evidence that female, minority and low-IAT engineers exhibit lower bias or discrimination in their code. However, we do find that prediction errors are correlated within demographic groups, which creates performance improvements through cross-demographic averaging. Finally, we quantify the benefits and tradeoffs of practical managerial or policy interventions such as technical advice, simple reminders, and improved incentives for decreasing algorithmic bias. Our full results are available at \href{https://ssrn.com/abstract=3615404}{https://ssrn.com/abstract=3615404}. 
\end{abstract}

  \defaultbibliography{bibliography}
  \defaultbibliographystyle{aer}
\begin{bibunit}

\emph{Our submission is a 4-page summary to comply with the workshop page limits. Our full results are available at \href{https://ssrn.com/abstract=3615404}{https://ssrn.com/abstract=3615404}.}

\section{Introduction}

Why do biased predictions arise, and what interventions can prevent them? Across a wide variety of theoretical models of behavior, biased predictions are responsible for demographic segregation and outcome disparities in settings including labor markets, criminal justice, and advertising. Although many classic theoretical models feature decision-makers with accurate (if discriminating) statistical predictors \citep{phelps1972statistical, arrow1973theory}, empirical evidence often shows that predictions are systematically inaccurate in practice \citep{bohren2019inaccurate}. How do these biased predictions arise? What theoretical mechanisms produce them, and what practical interventions can reduce prediction bias?

In this paper, we address these questions through a field experiment applying machine learning to predict workers’ performance. Automated hiring is a leading concern of policymakers questioning the ethics and fairness of AI systems. Research and public discourse on this topic have grown enormously in the past five years along with a growth in programs to introduce ethics into technical training. However, few studies have attempted to evaluate, audit, or learn from these interventions or connect them back to theory. This paper aims to step in that direction. 

We examine the formation of biased predictions in a unique field experiment about the development of AI technology. Our subjects are approximately 400 machine learning engineers. Our experiment gives us a direct view of prediction technology while it is being assembled. We then assess the resulting predictive algorithms using ground truth outcomes, and using randomized audit-like manipulations of algorithmic inputs. This setting creates measurement opportunities that would be impossible for learning processes in other settings, allowing us to study the mechanisms behind biased learning and prediction more directly. Through randomized treatments, we show how policy or managerial interventions can change the formation of biased predictions. 


\section{Empirical Setting}\label{EmpiricalSetting}

We conducted our experiment in two empirical settings. Approximately 80\% of our subjects were participants in a machine learning bootcamp at a large research university. The bootcamp taught machine learning programming techniques at a CS masters or advanced undergraduate level. We conducted our experiment in this setting twice; once in the Spring and once in the Fall versions of the bootcamp. 

Over half the bootcamp participants already graduated.\footnote{Approximately half had undergraduate degrees and half had MS or Ph.D. degrees (or pursuing them).} The average participant had 1.2 years of work experience (median 0.67 years) at the time of the experiment. These are attractive research subjects for our topic. Students from this program are frequently hired to work at large Internet companies such as Facebook and Google; the algorithms they will develop in the future may plausibly affect billions of Internet users. At the time of the experiment, 31\% of the bootcamp participants had already been employed by a household-name company as a software engineer. 

The programming task we studied was a graded homework assignment, and performance was highly incentivized and competitive. Although this setting has many attractive qualities for our experiment, it required some design adaptations to address the possibility of cheating and contamination between treatment groups (discussed in our full manuscript). 

To complement these relatively inexperienced subjects, we sought a population of programmers to complement the students: Freelance machine learning engineers. Using an online platform, we recruited 60 machine learning engineers (20\% of total subjects).\footnote{We created a private job listing on the platform, then randomly invited subjects from the machine learning section of the platform. We restricted our invitations to engineers who had at least 1.5 years of work experience, were based inside the United States, and whose rates were between \$20 and \$80.} The average contractor in our study was more experienced than our bootcamp programmers, worked 3.89 years before our experiment (median 3.16 years). 

Our full manuscript summarizes the characteristics of both sets of engineers in our experiment. Our engineers are 71\% male, 29\% female, 28\% White, 52\% East Asian, 15\% Indian, and only about 5\% Black or Latino/a/x. Like the broader population of engineers, our subject population lacks diversity in key characteristics. However, our sample is slightly more diverse than the US software engineering population as a whole.

\noindent {\bf Task}. All subjects in our experiment were assigned the same job: Develop an algorithm to predict math performance from biographical features on a job application, and apply it to 20,000 new individuals who do not appear in the training data. This task allows us to study a fundamental mechanism behind biased performance predictions.\footnote{There might be other biases in algorithms besides prediction bias which we do not discuss in this paper.} 

For reasons we discuss in our full paper, math is an attractive topic for empirical studies of algorithmic bias. As training data, engineers were given a sample of the OECD’s Programme for the International Assessment of Adult Competencies (``PIAAC'') dataset. PIAAC is the canonical dataset for cross-country and within-country comparisons of numeracy and skills. To facilitate others utilizing this dataset, we have made a cleaned and merged copy with documentation for other academic researchers.\footnote{Our cleaned version of this data ready for use by other researchers is available at \href{https://doi.org/10.7910/DVN/JAJ3CP}{https://doi.org/10.7910/DVN/JAJ3CP}.} The PIAAC data solves several critical research challenges for algorithmic bias researchers.

\section{Research Design}

Our paper features high-quality performance metrics about two groups of subjects. The first is our 400 AI programmers, and the second is the OECD test subjects. Performance for both groups can be measured objectively, avoiding the typical pitfalls of ``outcome tests'' such as subjective performance criteria. 

For the math subjects, performance on a standardized test is available for a professionally-weighted sample of the entire OECD population, and \emph{not} a limited, non-random subset. Regarding engineers, predictive outcomes are available for every engineer in our sample. For our engineering sample, we also have natural ways to aggregate individual performance to estimate team contributions. We can measure how correlated each engineer's model is with potential teammates, and how well a simple average (or more complicated aggregation) of two engineers perform as an ensemble. 

\noindent {\bf Experimental Treatment Arms}. Our research design included four main treatment arms. The first was a control group in which engineers were given PIAAC data featuring realistic sample selection problems in the training data, thus, we label this group as receiving ``biased training data.'' The second randomly-selected group received PIAAC data featuring no sample selection problems. In our remaining two treatments, engineers were given the first group’s training data (the biased training data, featuring sample selection problems). However, these experimental groups were also given policy interventions. The third group was given a non-technical reminder about the possibility of algorithmic bias. The fourth was given this reminder as well as a simple, jargon-free white paper about sample selection correction methods in machine learning \citep{zadrozny2003cost,chawla2002smote}.

{\bf Subtreatments: Performance Incentives}. Within each of the above treatments, subjects were randomized into varying performance incentive schemes. The goal of this randomization was to measure the effectiveness of using incentives to reduce algorithmic bias.

{\bf Audit Manipulations}. Our design also features an audit-like manipulation of algorithmic inputs. As part of our test data, we ask engineers to evaluate candidates whose characteristics are identical, except that a single covariate (gender) has changed. We can then compare the predicted outcomes of identical candidates, who differ only on their gender. Because the gender of the OECD subjects is accessible to the engineers’ programs, we can avoid the issues raised by manipulating first names. The outcomes of all predictions are far richer than a typical audit study; we estimate a continuous measure of predicted math performance that spans the entire spectrum.

Because our design features randomization both on screeners (programmers) and on candidates (subjects evaluated by the resume), our design resembles the two-sided audit \citep{agan2018ban, cowgill2019audit}. In these experiments, researchers hire professional recruiters to select resumes under experimental conditions. The recruiters are then asked to evaluate job applications with audit-like manipulations of names and characteristics. Our design is essentially identical, but we use \emph{software engineers} to build algorithms for decision-making, rather than human screeners.

{\bf Engineer Demographics}. Our subject population contained substantial variation in gender, race and other demographic characteristics. We utilized this diversity to measure whether demographically non-traditional programmers were more likely to notice and reduce algorithmic bias, and whether prediction errors were correlated within demographic groups.

\section{Results}

We find that biased predictions are mostly caused by biased training data. Having access to better training data helps engineers lower bias. One-third of the benefit of better training data comes through a novel economic mechanism: Engineers exert greater effort and are more responsive to incentives when given better training data. The group receiving better training data spent more hours working on their algorithms, and it was especially affected by our randomized manipulation of incentives. Our results suggest that higher quality data, incentives, and effort are complementary. With better data, the marginal benefit of working an additional hour is higher, and this leads programmers to exert additional effort. 


We also assess how performance varies with programmers' demographic characteristics, and with their performance on a psychological test of implicit bias (IAT) concerning gender and careers. We find no evidence that female, minority and low-IAT engineers exhibit lower bias or discrimination in their code. 

However, we do find that prediction errors are correlated within demographic groups, particularly gender. Specifically, two male programmers’ prediction errors are more likely to be correlated with each other. A team or ensemble approach that averages across two male programmers will effectively double down these errors. By contrast, a team that averages across gender introduces equally predictive but uncorrelated information. This creates performance improvements through cross-demographic averaging. 

Finally, we quantify the benefits and tradeoffs of practical managerial or policy interventions to decrease algorithmic bias. Our results suggest that simple reminders about bias are effective at improving algorithms’ accuracy, however, this result is mainly driven by a reduction of variance, and not through a reduction of bias. We find mixed results on our technical guidance interventions. Programmers who understand technical guidance successfully reduce bias. However, most do not follow the advice, resulting in algorithms that are worse than programmers given a simple reminder. 

\section{Conclusion}

As algorithms spread in influence, concern has grown about bias. However, the root causes of algorithmic bias are often unclear. In public discourse and academic literature, two theories of algorithmic bias have gained prominence. The first theory emphasizes ``Biased Training Data.'' Because machine learning applications are developed using historical data about outcomes, data coming from it would reflect and perpetuate any bias in the real world. A second theory emphasizes another factor ``Biased Programmers.'' This theory emphasizes the fact that programmers are highly non-representative and may exhibit biases  (consciously or otherwise) that are passed onto the algorithms they write. 

Both of these theories are likely contributors to algorithmic bias. However, the two theories require different policy solutions. In this paper, we seek to measure the relative contributions of biased data and biased programmers. 

Many of our results may be specific to our particular setting and interventions. Our paper should not be the final word on any of these topics, and there might be other sources of algorithmic bias besides prediction bias which we do not discuss in this paper. However, we do believe that empirical and experimental studies offer a novel, important and underutilized perspective on algorithmic bias.\footnote{Our approach is similar to ``empirical software engineering'' approaches \citep{shull2007guide, wohlin2012experimentation}, however, this field has been less likely to study algorithmic bias and fairness topics.} 

Questions about algorithmic bias are often framed as theoretical computer science problems. However, productionized algorithms are developed by humans, working inside organizations who are subject to training, persuasion, culture, incentives, and implementation frictions. An empirical, field experimental approach is also useful for evaluating practical policy solutions. As an example, the experiment above tests several plausible managerial and educational interventions for reducing algorithmic bias. 

The empirics of algorithmic bias also present unique opportunities for discrimination researchers. Algorithms are increasingly influential in core areas of discrimination including hiring, lending, and criminal justice. However, in addition, algorithms offer unique measurement opportunities that would be impossible in other economic settings. 

The ``source code,'' for human-driven decision-making -- or even a random sample of outputs -- is rarely available to any researcher. By contrast, algorithmic settings allowing inequality researchers to study the mechanisms behind discrimination more directly and transparently. This paper attempts to shed light on these processes -- and their relationship to realized prediction bias in algorithms -- using a large, preregistered randomized controlled trial. \qedsymbol






\putbib
\end{bibunit}
\end{document}